\documentclass[11pt,english]{article}
\usepackage{ae,aecompl}

\usepackage[]{fontenc}
\usepackage[latin9]{inputenc}
\usepackage{geometry}
\usepackage{color}
\usepackage{babel}
\usepackage{algorithm2e}
\usepackage{graphicx}
\inputencoding{utf8}
\usepackage[unicode=true,
 bookmarks=true,bookmarksnumbered=false,bookmarksopen=false,
 breaklinks=false,pdfborder={0 0 0},backref=false,colorlinks=true]
 {hyperref}
\hypersetup{pdftitle={ReconMap: An interactive visualisation of human metabolism},
 pdfauthor={Alberto Noronha, Anna DrÃ¶fn DaniÌelsdÃ³ttir, Freyr JÃ³hannsson , SoffiÌa JÃ³nsdÃ³ttir, Sindri Jarlssonâ¯ , JÃ³n PÃ©tur Gunnarsson, SigurÃ°ur BrynjÃ³lfsson, Piotr Gawron, Reinhard Schneider , Ines Thiele , and Ronan M. T. Fleming}}
\inputencoding{latin9}
\usepackage{breakurl}

\makeatletter
\newcommand{\lyxaddress}[1]{
\par {\raggedright #1
\vspace{1.4em}
\noindent\par}
}

\usepackage{mathtools,bbm}
\usepackage{bbold}
\usepackage[]{algorithm2e}
\usepackage{algpseudocode}

\algnewcommand\And{\textbf{and}}

\makeatother

\begin{document}

\title{ReconMap: An interactive visualisation of human metabolism}

\author{Alberto Noronha$^{1}$, Anna Dröfn Dan\'{i}elsdóttir$^{2}$, Freyr
Jóhannsson$^{2}$, Soff\'{i}a Jónsdóttir$^{2}$,\\
 Sindri Jarlsson\,$^{2}$, Jón Pétur Gunnarsson$^{2}$, Sigurdur
Brynjólfsson$^{2}$, Piotr Gawron$^{1}$,\\
 Reinhard Schneider$^{1}$, Ines Thiele$^{1}$, and Ronan M. T. Fleming$^{1}$\thanks{To whom correspondence should be addressed: ronan.mt.fleming@gmail.com}}

\maketitle

\lyxaddress{$^{1}$ Luxembourg Centre for Systems Biomedicine, University of
Luxembourg, Campus Belval, Esch-sur-Alzette, Luxembourg.\\
 $^{2}$ Center for Systems Biology, University of Iceland, Reykjavik,
Iceland.}
\begin{abstract}
A genome-scale reconstruction of human metabolism, Recon 2, is available
but no interface exists to interactively visualise its content integrated
with omics data and simulation results. We manually drew a comprehensive
map, ReconMap 2.0, that is consistent with the content of Recon 2.
We present it within a web interface that allows content query, visualization
of custom datasets and submission of feedback to manual curators.
ReconMap can be accessed via \href{http://vmh.uni.lu}{http://vmh.uni.lu},
with network export in a Systems Biology Graphical Notation compliant
format. A Constraint-Based Reconstruction and Analysis (COBRA) Toolbox
extension to interact with ReconMap is available via \href{https://github.com/opencobra/cobratoolbox}{https://github.com/opencobra/cobratoolbox}.
\end{abstract}

\section{INTRODUCTION}

A genome-scale metabolic reconstruction represents the full portfolio
of metabolic and transport reactions that can occur in a given organism.
From such reconstruction, a mathematical model can be derived, allowing
one to simulate of an organism's phenotypic behaviour under a particular
condition \cite{palsson334systems}. Recon 2 \cite{thiele2013community}
is a very comprehensive knowledge-base of human metabolism and has
been applied for numerous biomedical studies, including the mapping
and analysis of omics data sets \cite{aurich2016computational}. However,
despite numerous visualization efforts using automated layouts \cite{jensen2014metdraw},
there is no genome-scale yet biochemically intuitive human metabolic
map available for visualization of omic data in its network context.
Here, we release ReconMap, a comprehensive, manually curated map of
human metabolism presented utilising the Google Maps Application Programming
Interface (API) for highly responsive interactive navigation within
a platform that facilitates queries and custom data visualization.

\begin{figure*}[ht]
\centering \includegraphics[width=1\textwidth]{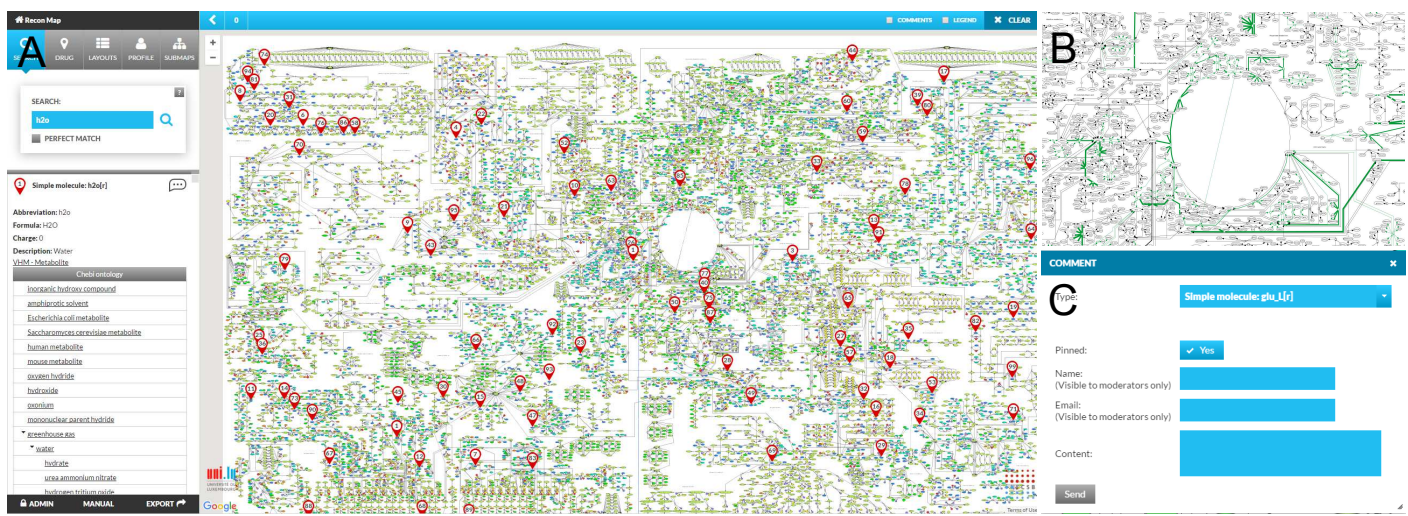} \caption{A - Web interface of ReconMap with search functionality. Information
retrieved for on a specific molecule are shown, along with external
liks; B - overlay of a flux distribution, using differential thickness
and color of the edges; C - Feedback interface that allows users to
provide suggestions and corrections to entities of the ReconMap and
Recon2.}
\label{fig:01} 
\end{figure*}

\section{FEATURES}

ReconMap content was derived from Recon 2.04, obtained from the Virtual
Metabolic Human database (VMH, \href{http://vmh.uni.lu}{http://vmh.uni.lu}).
Reactions (hyperedges) were manually laid out using the biochemical
network editor CellDesigner \cite{funahashi2008celldesigner}. Each
metabolite (node) was designated by its abbreviation and a letter
corresponding to the compartment, in which the reaction occurs (e.g.,
'{[}c{]}' for cytosol). Metabolites present in a high number of reactions
(e.g., common cofactors, water) were replicated across the map to
avoid excessive hyperedge crossover.

ReconMap is presented using the Molecular Interaction NEtwoRks VisuAlization
(MINERVA, \cite{gawron2016minerva}) platform built on the Google
Maps API, that together permit content query, low latency web display
and interactive navigation of generic and context-specific genome-scale
molecular networks. Each metabolite and reaction in ReconMap links
to the corresponding curated content provided by the VMH database.
Moreover, we also used MINERVA functions to connect to standard external
databases, such as the CHEMBL database \cite{bento2013chembl}.

\subsection{Overlay of simulation results and multi-omics datasets}

Recon-derived simulation results can be visualized on ReconMap using
a new extension to the COBRA Toolbox \cite{schellenberger2011quantitative}.
The user can perform a simulation, e.g., Flux Balance Anlalysis, using
the COBRA toolbox function 'optimizeCBmodel', then call the function
'buildFluxDistLayout' to write the input file for a ReconMap \textit{Overlay}.
This permits the user to translate each flux value into a custom thickness
and color within a simple tab-delimited highlight certain ReconMap
reactions. Similarly, registered users can display omic data on ReconMap
via the \textquotedbl{}Overlay\textquotedbl{} menu, which utilises
the MINERVA framework to generate a new ReconMap overlay that can
assign a different color and thickness to each node and reaction.

\subsection{Community-driven refinement of ReconMap \& Recon2}

All users may post suggestions for refinement and expansion that are
linked to a specific metabolite or reaction. Each suggestion is forwarded
to ReconMap and Recon2 curators for consideration when planning further
curation effort. As such, ReconMap enables the community-driven refinement
of human metabolic reconstruction and visualization.

\subsection{Connecting ReconMap and PDMap}

The Parkinson's disease map (PDMap, \cite{fujita2014integrating})
displays molecular interactions known to be involved in the pathogenesis
of Parkinson's disease. A total of 168 metabolites interlink ReconMap
and PDMap via standard identifiers. This feature is particularly interesting
when mapping omics datasets on both maps, thereby allowing the simultaneous
investigation of metabolic and non-metabolic pathways relevant for
Parkinson's and other neurodegenerative diseases.

\section{IMPLEMENTATION}

ReconMap was drawn using CellDesigner, is displayed using the MINERVA
platform, built on the Google Map API, using reconstruction content
from the VMH database \href{http://vmh.uni.lu}{http://vmh.uni.lu}.
Scripts for visualisation of COBRA Toolbox simulation results using
ReconMap are available here: \href{https://opencobra.github.io/}{https://opencobra.github.io/}.
To get started, see the tutorial here: \href{https://github.com/opencobra/cobratoolbox/tree/master/maps/ReconMap}{https://github.com/opencobra/cobratoolbox/tree/master/maps/ReconMap}.

\section{DISCUSSION}

ReconMap allows for efficient visualization of human metabolic reactions
and metabolites, connections with numerous online resources and the
VMH database, which hosts Recon 2 along with manually curated information.
ReconMap is a generic visualization of human metabolism and serves
as a template from which cell-, tissue-, and organ-specific maps may
be generated. Moreover, omics data can be directly mapped onto the
map as well as flux distributions resulting from simulations via an
extension to The COBRA Toolbox. ReconMap can be readily connected
to disease-specific maps, such as the Parkinson's disease map, thereby
enabling investigations beyond metabolic pathways. Future directions
include subsystem maps, conserved moiety tracing, drug target search,
and increased synergy with simulation tools.

\section*{ACKNOWLEDGEMENT}

\textit{Funding: }This work was supported by the Luxembourg National
Research Fund (FNR) through the National Centre of Excellence in Research
(NCER) on Parkinson's disease and the ATTRACT programme (FNR/A12/01),
and by the European Union's Horizon 2020 research and innovation programme
under grant agreement No 668738. \\
 \\
 Conflict of Interest: none declared.


\begin{thebibliography}{Schellenberger {it et~al}., 2011}
\bibitem[Aurich and Thiele, 2016]{aurich2016computational} Aurich,
M. and Thiele, I. (2016) Computational modeling of human metabolism
and its application to systems biomedicine, \textit{Systems Medicine},
253--281.

\bibitem[Bento {it et~al}., 2013]{bento2013chembl} Bento, A. \textit{et~al}.
(2013) The ChEMBL bioactivity database: an update, \textit{Nucleic
acids research}, gkt1031.

\bibitem[Fujita {it et~al}., 2014]{fujita2014integrating} Fujita,
K. \textit{et~al}. (2016) Integrating pathways of Parkinson's disease
in a molecular interaction map, \textit{Molecular neurobiology}, \textbf{49},
253--281.

\bibitem[Funahashi {it et~al}., 2008]{funahashi2008celldesigner}
Funahashi, A. \textit{et~al}. (2008) CellDesigner 3.5: a versatile
modeling tool for biochemical networks, \textit{Proceedings of the
IEEE}, \textbf{96}, 1254--1265.

\bibitem[Gawron {it et~al}., 2016]{gawron2016minerva} Gawron, P.
\textit{et~al}. (2016) MINERVA - a platform for visualization and
curation of molecular interaction networks, \textit{Submitted}.

\bibitem[Jensen and Papin, 2014]{jensen2014metdraw} Jensen, P. and
Papin, J. (2014) MetDraw: automated visualization of genome-scale
metabolic network reconstructions and high-throughput data, \textit{Bioinformatics},
\textbf{30}, 1327--1328.

\bibitem[Palsson, 2006]{palsson334systems} Palsson, B. \textit{et~al}.
(2006) Systems biology: properties of reconstructed networks, \textit{Cambridge
University Press}.

\bibitem[Schellenberger {it et~al}., 2011]{schellenberger2011quantitative}
Schellenberger, J. \textit{et~al}. (2011) Quantitative prediction
of cellular metabolism with constraint-based models: the COBRA Toolbox
v2. 0, \textit{Nature protocols}, \textbf{6}, 1290--1307.

\bibitem[Thiele {it et~al}., 2013]{thiele2013community} Thiele, I.
\textit{et~al}. (2013) A community-driven global reconstruction of
human metabolism, \textit{Nature biotechnology}, \textbf{31}, 419--425.\end{thebibliography}
\end{document}